\documentclass[12pt,a4paper]{article}
\usepackage[T2A]{fontenc}
\usepackage[utf8]{inputenc}
\usepackage[english]{babel}
\usepackage{amssymb}
\usepackage{amsmath}
\usepackage{graphicx}
\usepackage[small,sc,center]{titlesec}
\usepackage{indentfirst}
\usepackage{siunitx}
\usepackage[labelsep=period]{caption}
\usepackage{caption}

\newcommand{\nc}{\newcommand}
\nc{\etab}{\eta_\mathrm{B}}
\nc{\Menv}{M_\mathrm{env}}
\nc{\Teff}{T_\mathrm{eff}}
\nc{\tev}{t_\mathrm{ev}}

\addto\captionsenglish{}

\begin{document}

\begin{center}
	\textbf{Effects of metallicity on mode switching in Cepheids}
	
	\vskip 3mm
	\textbf{Yu. A. Fadeyev\footnote{E--mail: fadeyev@inasan.ru}}
	
	\textit{Institute of Astronomy, Russian Academy of Sciences,
		Pyatnitskaya ul. 48, Moscow, 119017 Russia} \\
	
	Received April 12, 2020; revised May 13, 2020; accepted May 13, 2020
\end{center}

\textbf{Abstract} ---
The mode switching in Cepheids is studied using the methods of the nonlinear theory of
stellar pulsation, depending on the main sequence mass $M_0$ and the abundance of elements
heavier than helium $Z$ .
The grid of evolutionary and hydrodynamic models of core--helium burning Cepheids is
represented by 30 evolutionary sequences of stars with initial masses
$5.7M_\odot\le M_0\le 7.2M_\odot$ and $Z=0.014$, 0.018, 0.022.
For considered values of $Z$ the periods of the fundamental mode and the first overtone at
the oscillation mode switching are shown to depend on the mean density of the stellar matter.
The upper limit of the period of the first overtone decreases witn increasing $Z$ from
$\approx 6.9$~day for $Z=0.014$ to $\approx 4.1$~day for $Z=0.022$.
The theoretical period--radius relation is independent of $Z$ and agrees well
(within 2.5\%) with recent measurements of Cepheid radii based on the Baade--Wesselink
method.
The fundamental parameters of the short--period Cepheid CG~Cas were derived with application
of observational estimates of the period and the rate of period change.
This star is shown to be the first--overtone pulsator.


\vskip 20mm
\section*{introduction}

At present the General Catalogue of Variable Stars (Samus' et al. 2017) contains the data on
nearly 640 Cepheids (i.e., $\delta$~Cep pulsating variables designated as DCEP and DCEPS).
Periods of light variations of these stars range from $\approx 2$ day to several dozen days.
All Cepheids with periods $\Pi > 7$~day are the fundamental mode pulsators, whereas variability
of many Cepheids with $\Pi < 7$~day  is due to instability of the first overtone.
Unfortunately, the particular threshold periods of the fundamental mode and the first overtone
corresponding to the mode switching in Cepheids do not exist as in the case of RR~Lyr type variables.
This is the cause of significant uncertainties arising in distinguishing the pulsation mode
of Cepheids with periods $5~\textrm{day}\lesssim\Pi\lesssim 7~\textrm{day}$.
The lack of the observational criterion for mode distinguishing leads to uncertainties in
the period--radius and period--luminosity relations because the fundamental mode and the first overtone
pulsators follow different relations
(Fernie 1968; B\"ohm--Vitense 1988, 1994; Sachkov 2002).
One of the possible solutions based on the Fourier decompostion of the Cepheid light curve was
proposed by Antonello and Poretti (1986) but it can only be applied in the case of high precision
photometric measurements.

The periods of the fundamental mode $\Pi_0^\star$ and the first overtone $\Pi_1^\star$
corresponding to oscillation mode switching are described with good accuracy as a function
of mean density $\bar\rho = M/(\frac{4}{3}\pi R^3)$, where $M$ and $R$ are the mass and
the radius of the Cepheid (Fadeyev 2020).
This conclusion is based on results of consistent stellar evolution and nonlinear stellar
pulsation calculations for Cepheid models with main sequence masses
$5.1M_\odot\le M_0\le 6.1M_\odot$ and the abundance of elements heavier than helium $Z=0.02$.
However recent estimates of the galactic abundance  gradient obtained from spectroscopic
observations of Cepheids are
$d[\textrm{Fe}/\textrm{H}]/dR_\textrm{G}\approx -0.06~\textrm{kpc}^{-1}$
and imply significant variation of the metal abundance $Z$ as a function
of galactocentric distance $R_\textrm{G}$ of the star
(Andrievsky et al. 2002; Lemasle et al. 2008; Luck and Lambert 2011; Luck et al. 2011.;
Genovali et al. 2014; Minniti et al. 2020).
Therefore, the dependencies $\Pi_0^\star(\bar\rho)$ and $\Pi_1^\star(\bar\rho)$
for other values of $Z$ remain uncertain.

The goal of the present study is to determine the dependencies
$\Pi_0^\star(\bar\rho)$ and $\Pi_1^\star(\bar\rho)$ for Cepheids with metal abundancies
$Z=0.014$, 0.018 and 0.022.
The initial conditions necessary for solution of the equations of radiation hydrodynamics
and time--dependent convection describing nonlinear stellar pulsations are obtained from
the grid of evolutionary sequences computed from the zero age main sequence to the stage
of helium exhaustion in the stellar core.
Evolutionary sequences were computed for stars with initial masses
$5.7M_\odot\le M_0\le 7.2M_\odot$ with the mass step $\Delta M_0 = 0.1M_\odot$.
The initial abundance of helium is assumed to be $Y=0.28$.

\section*{evolutionary sequences of cepheids}

As in the previous work of the author (Fadeyev 2020) the initial conditions for equations of radiation
hydrodynamics describing stellar pulsations were determined from selected stellar models of
evolutionary sequences computed with the program MESA version 12778 (Paxton et al. 2019).
Convective mixing was treated according to the standard theory (B\"ohm--Vitense 1958) with
mixing length to pressure scale height ratio $\alpha_\mathrm{MLT} = \Lambda/H_\mathrm{P} = 1.6$.
Extended convective mixing (overshooting) at convective boundaries was calculated using
the prescription of Herwig (2000) with parameters $f=0.016$ and $f_0=0.004$.
Effects of semiconvection were taken into account according to Langer et al. (1985) with efficiency
parameter $\alpha_\mathrm{sc}=0.1$ in the expression for the diffusion coefficient $D_\mathrm{sc}$.

To inhibit irregular growth of the convective core accompanied by appearance of spurious loops
in the HR--diagram the evolutionary calculations of the core helium burning stage  were carried out
with option 'conv\_premix\_avoid\_increase'.
The sufficiently small amplitude of central helium abundance jumps was reached owing to
significant increase of the mass zone number of the evolutionary model ($N\sim 4\times 10^4$)
together with reduction of the time step up to $\Delta t_\mathrm{ev} = 10^3$ yr.

The rate equations for nucleosynthesis were solved for the reaction network consisting of 26 isotopes
from hydrogen ${}^1\mathrm{H}$ to magnesium ${}^{24}\mathrm{Mg}$ with 81 reactions.
Calculation of reaction rates was carried out using the JINA Reaclib data (Cyburt et al. 2010).
Mass loss due to the stellar wind was computed by the Reimers (1975) formula with
parameter $\eta_\mathrm{R} = 0.3$.

\begin{figure}
\centerline{\includegraphics{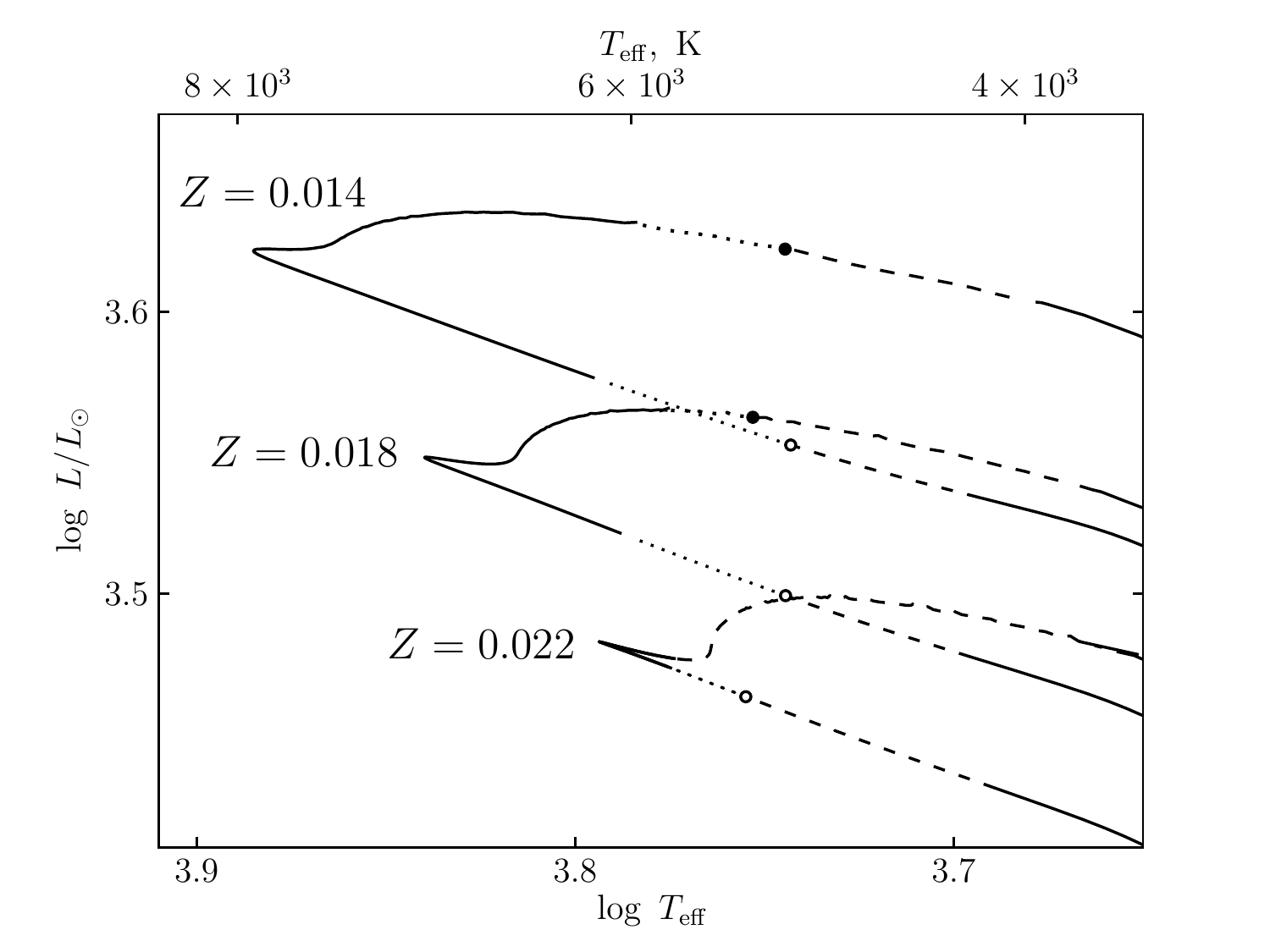}}
\caption{Evolutionary tracks of stars with initial mass $M_0=6.2M_\odot$ for metal abundances
         $Z=0.014$, 0.018 and 0.022 in vicinity of the Cepheid instability strip.
         Parts of the evolutionary tracks where the star does not pulsate are shown by solid lines.
         The dashed and the dotted lines indicate the stages when the Cepheid pulsates in the
         fundamental mode and in the first overtone, respectively.
         Open and filled circles correspond to the mode switching during the second and the third
         crossings of the instability strip.}
\label{fig1}
\end{figure}

The role of metallicity in evolution of Cepheids is illustrated in Fig.~\ref{fig1} where
the evolutionary tracks in vicinity of the second crossing and the third crossing
of the instability strip are shown in the HR diagram for stars with initial mass $M_0=6.2M_\odot$ and
metal abundances $Z=0.014$, 0.018, 0.022.
Parts of the evolutionary track with pulsations in the fundamental mode and the first overtone
are shown by dashed and dotted lines, respectively.

\begin{figure}
\centerline{\includegraphics{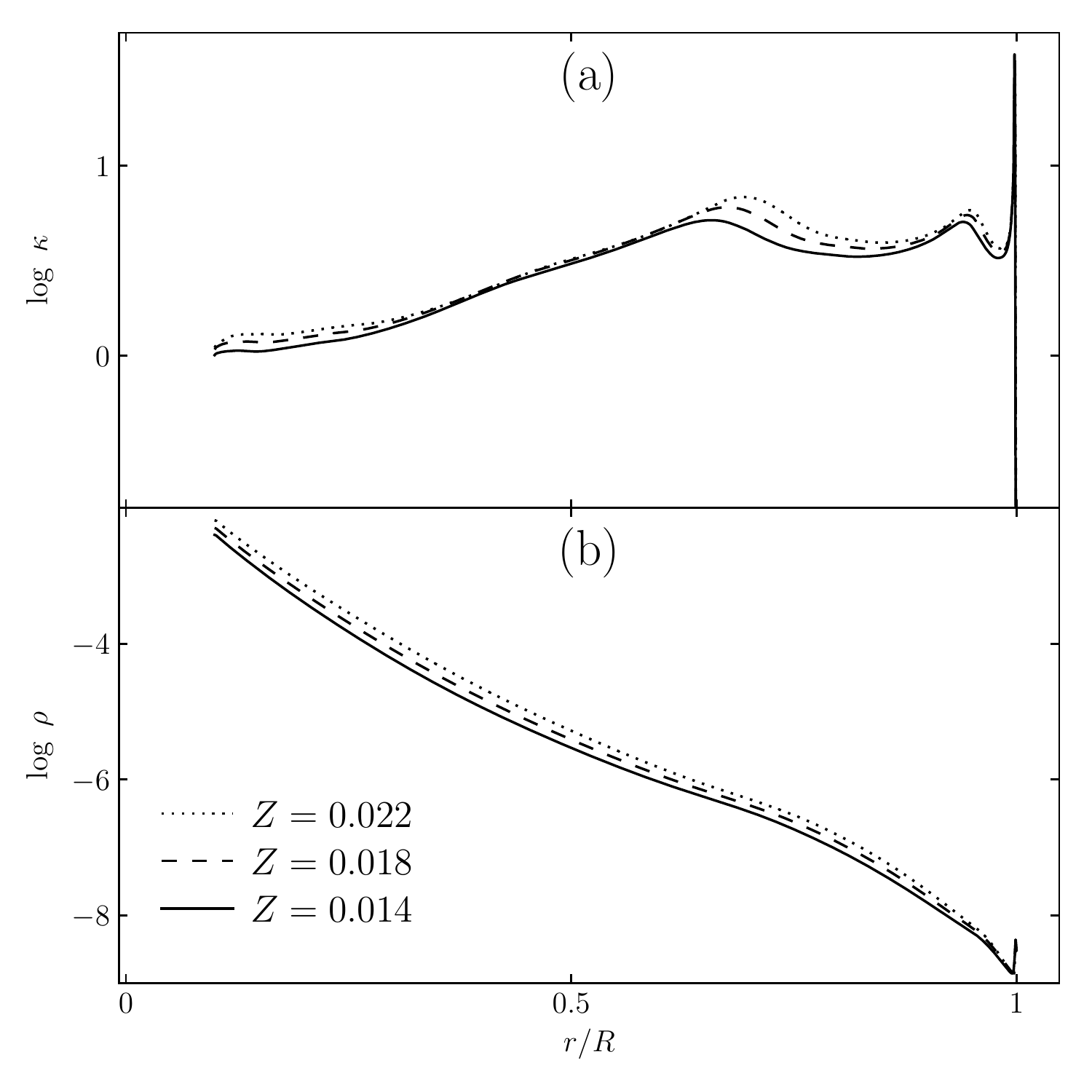}}
\caption{The opacity coefficient $\kappa$ (a) and the gas density $\rho$ (b) as a function of radial distance
         from the stellar center $r$ in hydrostatically equilibrium models at the oscillation mode
         swithing during the second crossing of the instability strip.
         Solid, dashed and dotted lines correspond to metal abundances $Z=0.014$,
         0.018 and 0.022. $R$ is the radius of the upper boundary of the evolutionary model.}
\label{fig2}
\end{figure}

Differences between evolutionary tracks shown in Fig.~\ref{fig1} are mainly due to the fact that
the opacity and the gas density in the envelope of the pulsating star increase with increasing $Z$.
The radial dependencies of the opacity $\kappa$ and the gas density $\rho$ are shown in Fig.~\ref{fig2}
for three hydrostatically equilibrium Cepheid models with different values of $Z$.
These models correspond to the mode switching during the second crossing of the instability strip
and are marked in Fig.~\ref{fig1} by open circles.

In Cepheid models considered in the present study the radius of the node of the first overtone
is $r_\mathrm{n}\approx 0.80R$, where $R$ is the radius of the upper boundary of the hydrostatically
equilibrium model.
The necessary condition for pulsations in the first overtone is that the helium ionizing zone
driving pulsation instability should locate in layers with radal distance $r > r_\mathrm{n}$.
This condition is not fulfilled for models of the evolutionary sequence $Z=0.022$, $M_0=6.2M_\odot$
during the third crossing of the instability strip so that the Cepheid evolves between the blue
and red edges of the instability strip as the fundamental mode pulsator (see Fig.~\ref{fig1}).

\section*{results of stellar pulsation calculations}

In this study we computed 30 evolutionary sequences for stars with initial metal abundances
$Z=0.014$, 0.018, 0.022 and initial masses $5.7M_\odot\le M_0\le 7.1M_\odot$.
Selected models of evolutionary sequences corresponding to stages of the second crossing and the third
crossing of the instability strip were used as initial conditions for solution of the equations
of radiation hydrodynamics and turbulent convection.
The equations of the problem are discussed in the earlier paper of the author (Fadeyev 2013).
Determination of the instability strip edges as well as calculation of the pulsation period of
the hydrpodynamic model and the star age at the mode switching is described by Fadeyev (2019).
The dependence $\Pi(\tev)$ was approximated by the algebraic polynom of the 3--rd order
for the evolutionary time interval with continuous change of the period.

\begin{figure}
\centerline{\includegraphics{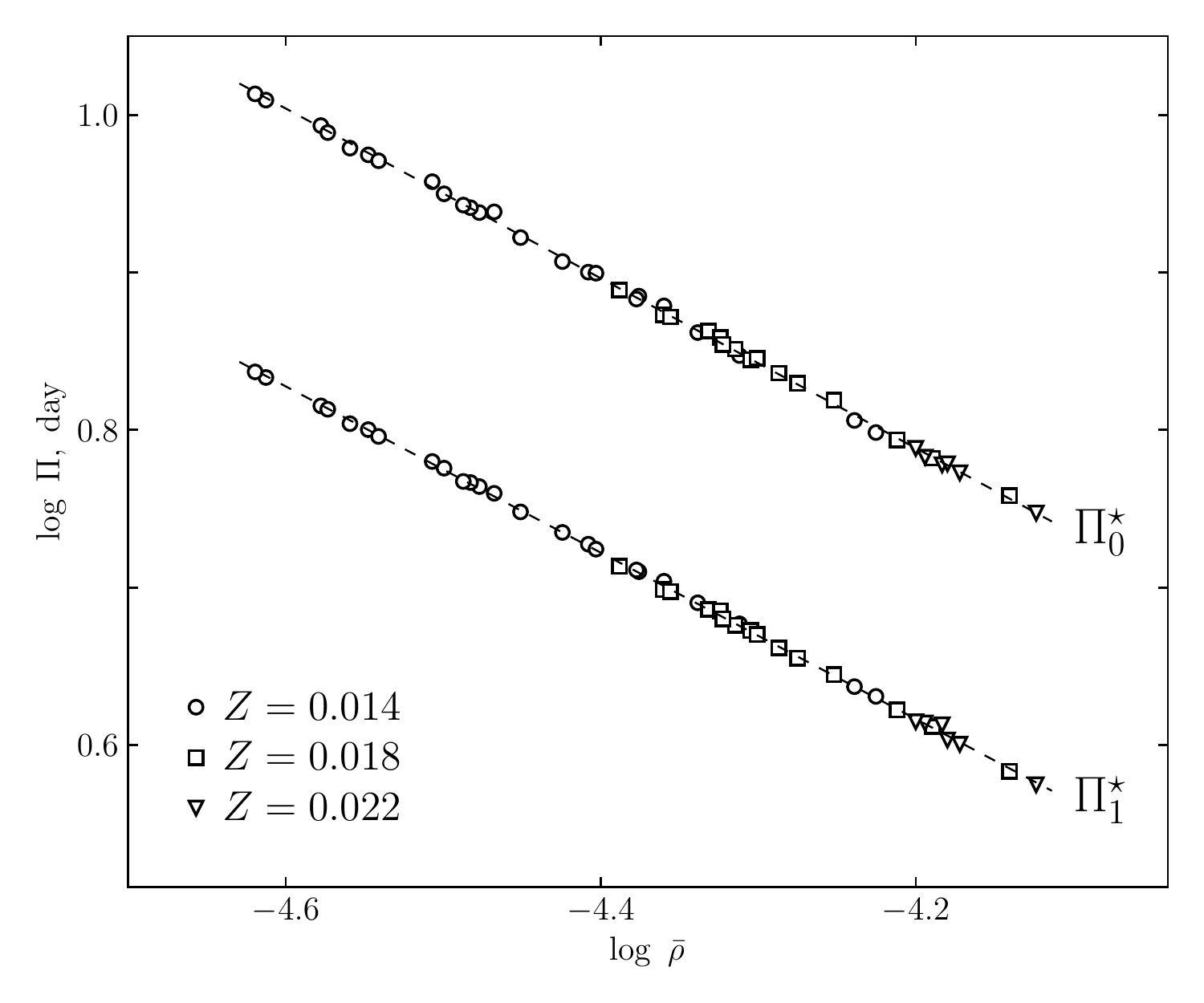}}
\caption{Periods of the fundamental mode $\Pi_0^\star$ and the first overtone $\Pi_1^\star$
         at the mode switching as a function of the mean density $\bar\rho$ for
         $Z=0.014$ (open circles), $Z=0.018$ (open squares) and $Z=0.022$ (open triangles).
         Dashed lines show approximations by relations (\ref{p0}) and (\ref{p1}).}
\label{fig3}
\end{figure}

The main goal of this work is to determine the periods of the fundamental mode $\Pi_0^\star$ and
the first overtone $\Pi_1^\star$ at the pulsation mode switching for evolutionary sequences with
specified values of $Z$ and $M_0$.
Results of calculations are shown in Fig.~\ref{fig3}, where periods $\Pi_0^\star$ and $\Pi_1^\star$
are plotted as a function of the mean stellar density $\bar\rho$.
As is seen, these dependencies are independent of $Z$ within the margin of error and are
given by relations
\begin{equation}
\label{p0}
\log\Pi_0^\star = -1.475 - 0.539 \log\bar\rho ,
\end{equation}
\begin{equation}
\label{p1}
\log\Pi_1^\star = -1.600   -0.528 \log\bar\rho .
\end{equation}

The plots in Fig.~\ref{fig3} illustrate the fact that periods of the fundamental mode and the first
overtone at the mode switching increase with increasing mass and increasing radius of the Cepheid
(i.e. with decreasing $\bar\rho$) irrespective of $Z$.
Decrease of the stellar mean density is accompanied by extension of the helium ionizing zone so that
pulsations in the first overtone become impossible once the lower boundary of the ionizing zone
reaches the node of the first overtone.
As seen in Fig.~\ref{fig3}, the interval of existence of oscillations in the first overtone
(i.e. with period $\Pi_1 < \Pi_1^\star$) reduces with increasing $Z$.

For the sake of clarity, properties of the models at the upper limit of existence of
the first overtone oscillations are listed in Table~\ref{tabl1}.
First three columns of Table~\ref{tabl1} give the metal abundandance $Z$, initial stellar mass $M_0$ and
the number of the instability strip crossing $i$.
The fourth and the fifth columns give the stellar mass $M$ and the mean density $\bar\rho$
at the mode switching.
Last two columns give the pulsation periods $\Pi_0^\star$ and $\Pi_1^\star$.
It should be noted that during the second crossing of the instability strip ($i=2$)
the radial pulsations switch from the fundamental mode to first overtone, whereas during the third
crossing ($i=3$) pulsations switch from the first overtone to the fundamental mode.

\section*{period--radius relation}

Measurements of Cepheid radii using the Baade--Wesselink method (Baade, 1926; Wesselink, 1946)
provide the basis for the period--radius relation of galactic Cepheids which is applied for
calibration of the period--luminosity relation 
(Kervella et al. 2004; Turner 2010; Molinaro et al. 2011; Lazovik and Rastorguev 2020).
The period--radius relation is derived for fundamental mode pulsators and the Cepheids
pulsating in the first overtone are included by multiplication of the period by the constant
factor.
For example, Lazovik and Rastorguev (2020) used the factor 1.41 to this end.

On the whole, in the present work we computed 326 hydrodynamic models of Cepheids, where 193
models are the fundamental mode pulsators and 133 models are the first overtone pulsators.
The pulsation period of the hydrodynamic model was calculated using the discrete Fourier
transform of the kinetic energy of pulsation motions (Fadeyev 2013).
Together with period of the principal mode we evaluated also the period of the secondary mode.
In the fundamental mode pulsator the role of the secondary mode is played by the first
overtone, whereas in the case of the first overtone pulsator the role of the secondary mode
belongs to the fundamental mode.
The period ratio of the fundamental mode and the first overtone was found to be
$\Pi_0/\Pi_1\approx 1.50$ and independent of the locus with respect to the mode switching.

\begin{figure}
\centerline{\includegraphics{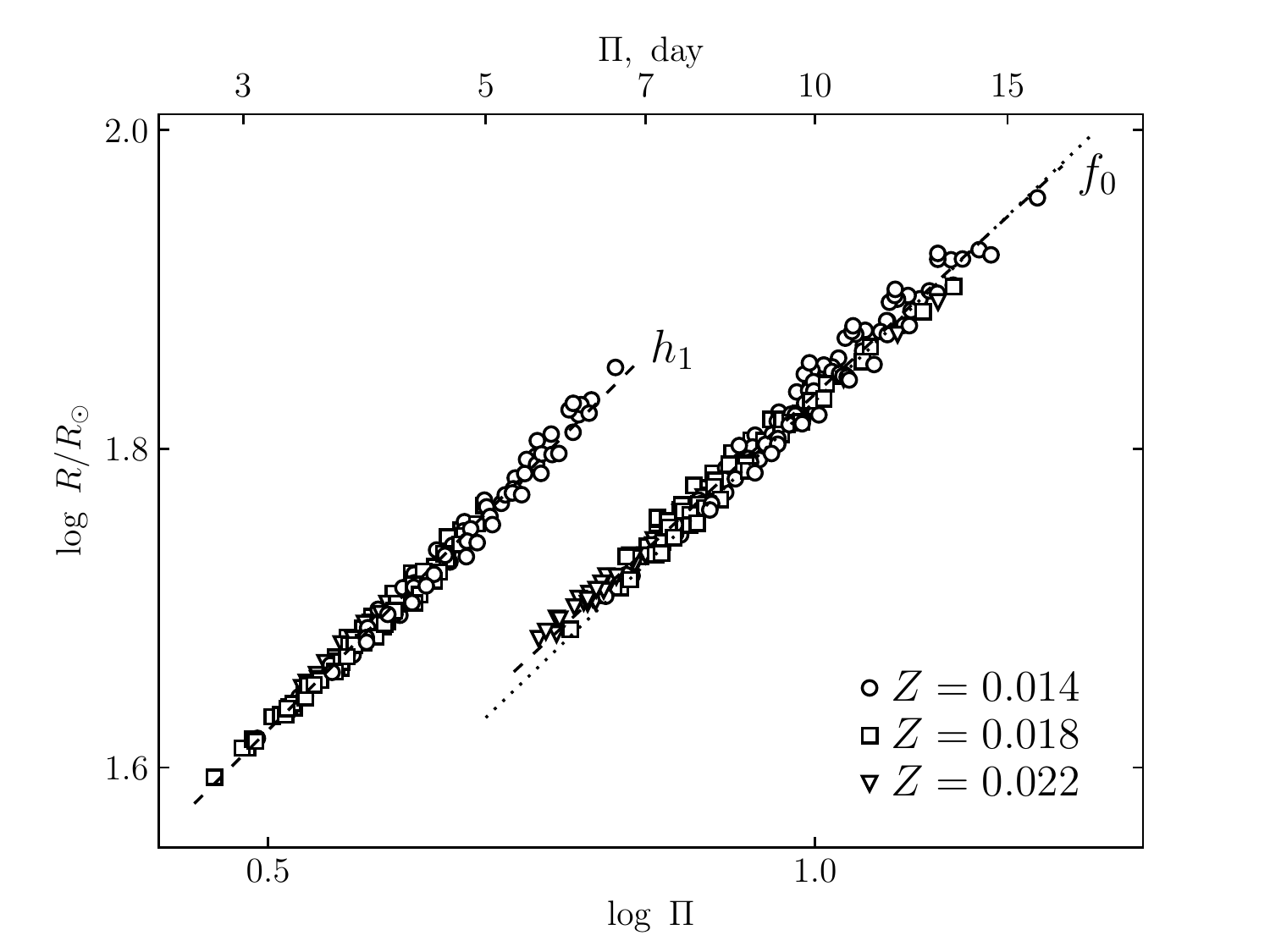}}
\caption{Period--radius relations obtained from hydrodynamic computations for Cepheids pulsating
         in the fundamental mode ($f_0$) and in the first overtone ($h_1$).
         Dashed lines represent relations (\ref{f0}) and (\ref{h1}).
         Period--radius relation (\ref{lr20}) from Lazovik and Rastorguev (2020)
         is shown by the dotted line.}
\label{fig4}
\end{figure}

The period--radius relations obtained from the present evolutionary and hydrodynamic calculations
are shown in Fig.~\ref{fig4}.
As in Fig.~\ref{fig3}, the plots in Fig.~\ref{fig4} do not show any dependence on $Z$,
so that the periods and radii locate near the straigth lines $f_0$ and $h_1$ which are
given by
\begin{equation}
\label{f0}
 \log R/R_\odot = 1.203 + 0.631 \log\Pi_0
\end{equation}
and
\begin{equation}
\label{h1}
 \log R/R_\odot = 1.281 + 0.685 \log\Pi_1 ,
\end{equation}
where the pulsation period is expressed in days.

For hydrodynamic models considered in the present study the lower limit of the fundamental mode period
is $\Pi_0=5.59$~day, whereas the upper limit of the first overtone is $\Pi_1=6.57$~day.
In the range of these two values the fundamental mode to first overtone period ratio
evaluated from (\ref{f0}) and (\ref{h1}) varies within $1.49\le\Pi_0/\Pi_1\le 1.56$.
As seen in Fig.~\ref{fig4}, the scatter of period ratios around $\Pi_0/\Pi_1=1.50$ is due to
variation of radii near the regression line.

In recent years different realizations of the Baade--Wesselink technique have been developed
in order to improve the accuracy of Cepheid radius measurements, so that at present there is
a number of empirical period--radius relations.
These methods were briefly reviewed by Lazovik and Rastorguev (2020).
In the present work the theoretical period--radius relation (\ref{f0}) was compared with
empirical relations derived by 
Sachkov et al. (1998), Turner and Burke (2002), Kervella et al. (2004), Storm et al. (2004),
Groenewegen (2007), Molinaro et al. (2011), Gallenne et al. (2017), Lazovik and Rastorguev (2020).
The best agreement of the theoretical relation (\ref{f0}) with empirical period--radius
relation was found for
\begin{equation}
\label{lr20}
\log R/R_\odot = 1.17 + 0.66 \log\Pi_0
\end{equation}
derived by Lazovik and Rastorguev (2020).
In Fig.~\ref{fig4} relation (\ref{lr20}) is shown by the dotted line.
For fundamental mode pulsators with period $\Pi=5.6$~day the difference between
(\ref{f0}) and (\ref{lr20}) does not exceed $\approx 2.6\%$,
whereas for periods $\Pi\approx 15$~day the difference between the theory and observations
becomes as small as $\approx 0.2\%$, so that the theoretical and empirical plots become almost
the same.

\section*{fundamental parameters of the cepheid cg cas}

The short--period Cepheid CG~Cas ($\Pi=4.3656$~day) is observed in the corona of the young open cluster
Berkeley~58 and is of great interest for calibration of the period--luminosity relation.
The $O-C$ diagram of CG~Cas spans somewhat more than a hundred years and is fitted by the parabolic
dependence indicating the secular period growth with rate $\dot\Pi=0.170$~s/yr (Turner et al. 2008).
Thus, there is an opportunity to evaluate the fundamental parameters of this star on the basis
of consistent stellar evolution and nonlinear stellar pulsation calculations.
However, the earlier attempt to determine the fundamental parameters of CG~Cas for the evolutionary
sequences with metal abundance $Z=0.02$ failed because the upper limit of the first overtone period
was found to be shorter than the period of CG~Cas (Fadeyev 2020).

\begin{figure}
\centerline{\includegraphics{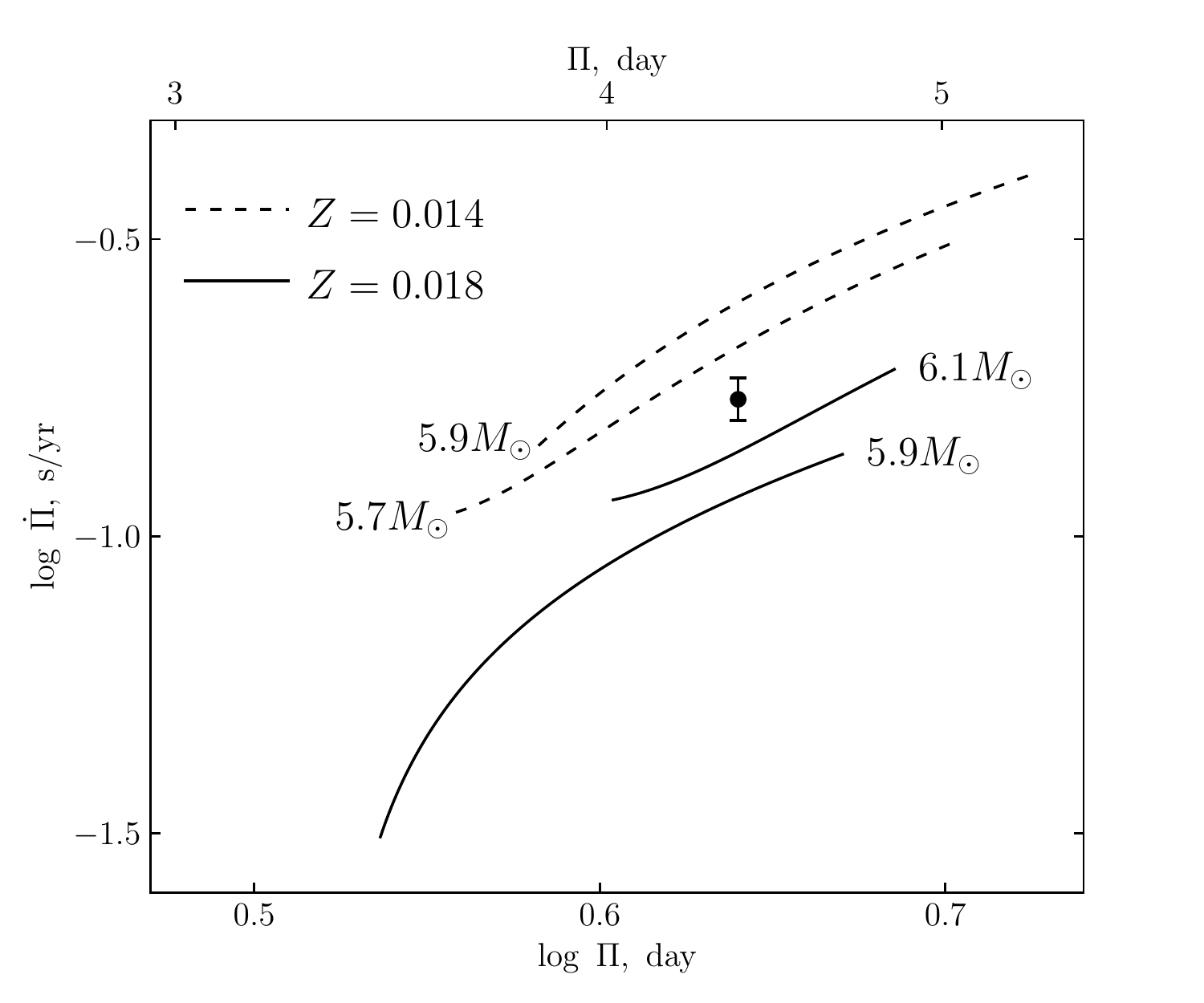}}
\caption{The diagram $\Pi-\dot\Pi$ for Cepheid models pulsating in the first overtone
         during the third crpossing of the instability strip for $Z=0.014$ (dashed lines)
         and $Z=0.018$ (solid lines). The initial mass $M_0$ is indicated at the curves.
         Observational estimates of $\Pi$ и $\dot\Pi$ for CG~Cas (Turner et al. 2008) are
         shown by the filled circle.}
\label{fig5}
\end{figure}

As seen in Fig.~\ref{fig3} and \ref{fig4}, for evolutionary sequences with metal abundances
$0.014\le Z\le 0.022$ the radial pulsations with period $\Pi=4.3656$~day in the fundamental mode
should be excluded since the lower limit of the fundamental mode period is $\Pi=5.59$~day.
Thus, to reconcile the theory with observations we have to admit that CG~Cas is the first overtone
pulsator.
The plots of the period change rate as a function of the first overtone period during the third
crossing of the instability strip are shown in Fig.~\ref{fig5} for several evolutionary sequences
with $Z=0.014$ and $Z=0.018$.
Each plot displays the evolution of $\Pi$ and $\dot\Pi$ between the blue edge of the
instability strip and the mode switching from the first overtone to the fundamental mode.

As seen in Fig.~\ref{fig5}, the evolutionary sequences $Z=0.014$, $M_0=5.7M_\odot$ and
$Z=0.018$, $M_0=6.1M_\odot$ are the nearest to the observational estimates of $\Pi$ and $\dot\Pi$.
Substitution of $\Pi=4.3656$~day into dependencies $\Pi(\tev)$ of these evolutionary sequences
yields the star age $\tev$ and, therefore, its mass, radius and luminosity.
The fundamental parameters of the Cepheid CG~Cas are listed in Table~\ref{tabl2}, where
in the last column we give the metallicity index of the evolutionary sequence evaluated for the
solar metal abundance $Z_\odot=0.0134$ (Asplund et al. 2009).
The metallicity index of CG~Cas is $[\textrm{Fe}/\textrm{H}]=0.09$ (Genovali et al. 2014)
therefore the model of the evolutionary sequence $Z=0.018$, $M_0=6.1M_\odot$ seems to be
the most preferable.

Bearing in mind the results presented above we have to note that in the works devoted to
calibration of the period--luminosity relation on the basis of Gaia trigonometric
parallaxes (Ripepi et al. 2019; Breuval et al.  2020)
the Cepheid CG~Cas is improperly considered as a fundamental mode pulsator.

\section*{conclusions}

Dependence of the stellar opacity on $Z$ significantly affects the gas density in the stellar envelope
and location of the pulsation driving zone with respect to the node of the first overtone.
Increase of the mass and the radius of the Cepheid is accompanied by decreasing mean density
and increasing width of the helium ionizing zone responsible for excitation of oscillations.
Pulsations in the first overtone become impossible when the inner boundary of the helium ionizing
zone reaches the node of the first overtone during pulsation motions.
Increase of periods $\Pi_0^\star$ and $\Pi_1^\star$ corresponding to the mode switching
is due to decrease of the gas density in the envelope of the pulsating star.
Existence of the upper limit $\Pi_1^\star$ implies that the necessary condition for
oscillations in the first overtone is not fulfilled for $\Pi_1 > \Pi_1^\star$.

The lower period limit of radial pulsations in the fundamental mode is $\Pi_0^\star=5.59$ day
for models of the evolutionary sequence $Z=0.022$, $M_0=5.9M_\odot$, whereas the upper period limit
of oscillations in the first overtone is $\Pi_1^\star=6.87$ day for $Z=0.014$, $M_0=7.1$.
In other words, for metal abundances $0.014\le Z\le 0.022$ Cepheids can oscillate in the
fundamental mode with periods $\Pi>5.6$ day, whereas oscillations in the first overtone are possible
with periods $\Pi<6.9$ day.
However the index of metallicity of the most of Cepheids with accurate measurements of
the iron abundance decrease from $[\textrm{Fe/H}]$ = 0.27 at the galactocentric distance
$R_\mathrm{G}=5$ kpc to $[\textrm{Fe/H}]$ = 0.27 at $R_\mathrm{G}=17$ kpc (Genovali et al. 2014).
Adopting the solar galactocentric distance to be $R_{\mathrm{G}\odot}=7.94$~kpc
(Groenewegen et al. 2008) and the solar metal abundance $Z_\odot = 0.0134$ 
(Asplund et al.  2009) we find that in the present work we considered the models with metallicity
indices $0.012\le [\textrm{Fe/H}]\le 0.215$.
Therefore, the Cepheid models computed in the present study are mostly confined to stars located
between the Sun and the center of Galaxy, albeit this region can be somewhat wider due to
natural dispersion of metallicity in Cepheids.
Extention of the Cepheid grid to lower $Z$ will allow us to consider the stars with galactocentric
distances greater that the solar galactocentric distance and may lead to somewhat larger values
of $\Pi_1^\star$.
Enlargerement of $\Pi_1^\star$ due to decrease of $Z$ can be evaluated from additional evolutionary
and hydrodynamic computations.

\newpage
\section*{references}

\begin{enumerate}

\item S.M. Andrievsky, V.V. Kovtyukh, R.E. Luck, J.R.D. L\'epine,
      D. Bersier, W.J. Maciel, B. Barbuy, V.G. Klochkova, V.E. Panchuk, and R.U. Karpischek,
      Astron. Astrophys. \textbf{381}, 32 (2002).

\item E. Antonello and E. Poretti, Astron. Astrophys. \textbf{169}, 149 (1986).

\item M. Asplund, N. Grevesse, A.J. Sauval, and P. Scott,
      Annual Rev. Astron. Astrophys. \textbf{47}, 481 (2009).

\item W. Baade, Astronomische Nachrichten \textbf{228}, 359 (1926).

\item E. B\"ohm--Vitense, Zeitschrift f\"ur Astrophys. \textbf{46}, 108 (1958).

\item E. B\"ohm--Vitense, Astrophys. J. \textbf{324}, L27 (1988).
 
\item E. B\"ohm--Vitense, Astron. J. \textbf{107}, 673 (1994).

\item L. Breuval, P. Kervella, R.I. Anderson, A.G. Riess, F. Arenou, B. Trahin, A. M\'erand,
      A. Gallenne, W. Gieren, J. Storm, G. Bono, G. Pietrzy\'nski, N. Nardetto, B. Javanmardi, V. Hocd\'e,
      Astron. Astrophys. \textbf{643}, A115 (2020).

\item R.H. Cyburt, A.M. Amthor, R. Ferguson, Z. Meisel, K. Smith,
      S. Warren, A. Heger, R.D. Hoffman, T. Rauscher, A. Sakharuk, H. Schatz,
      F.K. Thielemann, and M. Wiescher,
      Astrophys. J. Suppl. Ser. \textbf{189}, 240 (2010).

\item Yu.A. Fadeyev, Astron. Lett. \textbf{39}, 306 (2013).

\item Yu.A. Fadeyev, Astron. Lett. \textbf{45}, 353 (2019).

\item Yu.A. Fadeyev, Astron. Lett. \textbf{46}, 324 (2020).

\item J.D. Fernie, Astrophys. J. \textbf{151}, 197 (1968).

\item A. Gallenne, P. Kervella, A. M\'trand, G. Pietrzy\'nski, W. Gieren, N. Nardetto, and B. Trahin,
      Astron. Astrophys. \textbf{608}, A18 (2017).
      
\item K. Genovali, B. Lemasle, G. Bono, M. Romaniello, M. Fabrizio, I. Ferraro,
      G. Iannicola, C.D. Laney, M. Nonino, M. Bergemann, R. Buonanno, P. Fran\c cois, 
      L. Inno, R.--P. Kudritzki, N. Matsunaga, S. Pedicelli, F. Primas, and F. Th\' evenin,
      Astron. Astrophys. \textbf{566}, A37 (2014).

\item M.A.T. Groenewegen, Astron. Astrophys. \textbf{474}, 975 (2007).

\item M.A.T. Groenewegen, A. Udalski, and G. Bono, Astron. Astrophys. \textbf{481}, 441 (2008).

\item F. Herwig, Astron. Astrophys. \textbf{360}, 952 (2000).

\item P. Kervella, D. Bersier, D. Mourard, N. Nardetto, and V. Coud\'e du Foresto,
      Astron. Astrophys. \textbf{423}, 327 (2004).

\item N. Langer, M.F. El Eid, and K.J. Fricke, Astron. Astrophys. \textbf{145}, 179 (1985).

\item Ya.A. Lazovik and A.S. Rastorguev, Astron. J. \textbf{160}, 136 (2020).

\item B. Lemasle, P. Fran\c cois, A. Piersimoni, S. Pedicelli, G. Bono,
     C.D. Laney, F. Primas, and M. Romaniello, Astron. Astrophys. \textbf{490}, 613 (2008).
      
\item R.E. Luck and D.L. Lambert, Astron. J. \textbf{142}, 136 (2011).

\item R.E. Luck, S.M. Andrievsky, V.V. Kovtyukh, W. Gieren, and D. Graczyk,
      Astron. J. \textbf{142}, 51 (2011).

\item J.H. Minniti, L. Sbordone, A. Rojas--Arriagada, M. Zoccali, R. Contreras Ramos,
      D. Minniti, M. Marconi, M.; V.F. Braga, M. Catelan, S. Duffau, W. Gieren, W.; and
      A.A.R. Valcarce, Astron. Astrophys. \textbf{640}, A92 (2020).

\item R. Molinaro, V. Ripepi, M. Marconi, G. Bono, J. Lub, S. Pedicelli, and J.W. Pel,
      Mon. Not. R. Astron. Soc. \textbf{413}, 942 (2011).

\item B. Paxton, R. Smolec, J. Schwab, A. Gautschy, L. Bildsten, M. Cantiello, A. Dotter,
      R. Farmer, J.A. Goldberg, A.S. Jermyn, S.M. Kanbur, P. Marchant, A. Thoul,
      R.H.D. Townsend, W.M. Wolf, M. Zhang, and F.X. Timmes,
      Astrophys. J. Suppl. Ser. \textbf{243}, 10 (2019).

\item D. Reimers, \textit{Problems in stellar atmospheres and envelopes}
      (Ed. B. Baschek, W.H. Kegel, G. Traving, New York: Springer-Verlag, 1975), p. 229.

\item V. Ripepi, R. Molinaro, I. Musella, M. Marconi, S. Leccia, and L. Eyer,
      Astron. Astrophys. \textbf{625}, A14 (2019).

\item M.E. Sachkov, Astron. Lett. \textbf{28}, 589 (2002).

\item M.E. Sachkov, A.S. Rastorguev, N.N. Samus', and N.A. Gorynya, Astron. Lett. \textbf{24}, 377 (1998).

\item N.N. Samus', E.V. Kazarovets, O.V. Durlevich, N.N. Kireeva, and E.N. Pastukhova,
      Astron. Rep. \textbf{61}, 80 (2017).

\item J. Storm, B.W. Carney, W.P. Gieren, P. Fouqu\'e, D.W. Latham, and A.M. Fry,
      Astron. Astrophys. \textbf{415}, 531 (2004).

\item D.G. Turner, Astrophys. Space Sci. \textbf{326}, 219 (2010).

\item D.G. Turner and J.F. Burke, Astron. J. \textbf{124}, 2931 (2002).

\item D.G. Turner, D. Forbes, D. English, P.J.T. Leonard, J.N. Scrimger,
      A.W. Wehlau, R.L. Phelps, L.N. Berdnikov, and E.N. Pastukhova,
      MNRAS \textbf{388}, 444 (2008).

\item A.J. Wesselink, Bulletin of the Astronomical Institutes of the Netherlands \textbf{10}, 91 (1946).

\end{enumerate}

\newpage
\begin{table}
\caption{Upper limits of the periods of the fundamental mode $\Pi_0^\star$ and
         the first overtone $\Pi_1^\star$ at the mode switching}
\label{tabl1}
\begin{center}
\begin{tabular}{c|c|c|c|c|c|c}
\hline
 $Z$ & $M_0/M_\odot$ & $i$ & $M/M_\odot$ & $\lg\bar{\rho}$ & $\Pi_0^\star, \textrm{day}$ & $\Pi_1^\star, \textrm{day}$ \\
\hline
 0.014 & 7.1 & 2 & 7.055 & -4.620 & 10.312 &  6.868 \\
       & 6.5 & 3 & 6.451 & -4.578 &  9.844 &  6.536 \\
 0.018 & 6.5 & 2 & 6.469 & -4.388 &  7.741 &  5.170 \\
       & 6.2 & 3 & 6.162 & -4.360 &  7.467 &  4.997 \\
 0.022 & 6.2 & 2 & 6.175 & -4.200 &  6.142 &  4.118 \\
       & 6.1 & 3 & 6.074 & -4.194 &  6.063 &  4.111 \\
\hline
\end{tabular}
\end{center}
\end{table}

\begin{table}
\caption{Fundamental parameters of the Cepheid CG~Cas}
\label{tabl2}
\begin{center}
\begin{tabular}{c|c|c|c|c|c|c|c}
\hline
 $Z$ & $M_0/M_\odot$ & $\tev,\ 10^6\textrm{yr}$ & $M/M_\odot$ & $L/L_\odot$ & $R/R_\odot$ & $T_\textrm{eff}$, K & $[\textrm{Fe}/\textrm{H}]$\\
\hline
0.014 & 5.7 & 76.217 &   5.664 &  3075 &  52.16 &  5955 & 0.019 \\
0.018 & 6.1 & 67.256 &   6.065 &  3429 &  53.39 &  6049 & 0.128 \\
\hline
\end{tabular}
\end{center}
\end{table}
\clearpage

\newpage
\clearpage

\newpage
\clearpage

\newpage
\clearpage

\newpage
\clearpage

\newpage
\clearpage

\end{document}